\documentclass[12pt,a4paper,preprintnumbers,superscriptaddress]{revtex4-1} %%% Use this class to prepare your letter for review
\usepackage[fleqn]{amsmath}
\usepackage{amssymb}
\usepackage{bm}% bold math
\usepackage{graphicx} % \includegraphics
\usepackage{subfigure}
\usepackage{mathrsfs}
\usepackage{color}
\usepackage[pdfstartview=FitH,
            CJKbookmarks=true,
            bookmarksnumbered=true,
            bookmarksopen=true,
            colorlinks=true,
            pdfborder=001,
            linkcolor=blue,
            citecolor=blue,
            urlcolor=blue
            ]{hyperref}
\RequirePackage{color}
\begin{document}

\title{Short-time evolution of Lagrangian velocity gradient correlations in isotropic turbulence}

\author{L. Fang}
\affiliation{LMP, Ecole Centrale de P\'{e}kin, Beijing University of Aeronautics and Astronautics, Beijing 100191, China}

\author{W.J.T. Bos}
\affiliation{LMFA, CNRS, Ecole Centrale de Lyon - Universit\'{e} de Lyon, 69130 Ecully, France}

\author{G.D. Jin}
\thanks{Corresponding Author. Email: gdjin@lnm.imech.ac.cn}
\affiliation{LNM, Institute of Mechanics, Chinese Academy of Sciences, Beijing 100190, China}

\date{\today}

\begin{abstract}
\noindent \textbf{Abstract} We show by direct numerical simulation (DNS) that the Lagrangian cross correlation of velocity gradients in homogeneous isotropic turbulence increases at short times, whereas its auto-correlation decreases. Kinematic considerations allow to show that two invariants of the turbulent velocity field determine the short-time velocity gradient correlations. In order to get a more intuitive understanding of the dynamics for longer times, heuristic models are proposed involving the combined action of local shear and rotation. These models quantitatively reproduce the effects and disentangle the different physical mechanisms leading to the observations in the DNS.
\end{abstract}

%Use showkeys class option if keyword display desired

\maketitle

\section{Introduction}
The study of the Lagrangian dynamics of the velocity gradient tensor in turbulent flows has recently received a considerable amount of attention. Its investigation helps us to understand important phenomena involving the small scales of turbulent flows, such as the preferential vorticity alignment, the skewness of longitudinal velocity gradients, small scale intermittency, \textit{etc.} \cite{Meneveau2011}. Recent developments in experimental techniques  allow nowadays to deterministically track the development of the velocity gradient tensor in Eulerian \cite{wallace2010measurement} and Lagrangian settings \cite{guala2007experimental}. These results, combined with  results from Direct Numerical Simulations (DNS) (\textit{e.g.} \cite{ashurst1987alignment,ishihara2007small}) provide unprecedented possibilities to understand the origin of these interesting phenomena.

In order to understand the phenomenology of the velocity gradient evolution, simplified models are needed which allow to identify the different physical features. One of the simplest models is the so-called restricted Euler (RE) equation for the velocity gradient tensor. This equation is obtained by removing the viscous diffusion term and the anisotropic part of the pressure Hessian from the evolution equation of the velocity-gradient tensor, only leaving the advection term, the self-stretching term and the symmetric part of the pressure Hessian~\cite{Cantwell1992}. There are many studies on the properties of this RE system, investigating in particular the evolution of the invariants and their probability density functions \cite{Cantwell1992,Cantwell1993, Martin1998, Li2005}. However, the system evolves to a singular state \cite{Cantwell2002}, so that the time-evolution at long times is not comparable to realistic Navier-Stokes dynamics.  In particular, the influence of the anisotropic part of the pressure
Hessian on the distribution of tensor invariants seems to be significant~\cite{Ooi1999, Luthi2009}.  More recent theoretical approaches attempt to model these remaining terms using geometrical considerations~\cite{chertkov1999lagrangian,naso2005scale}, assumptions on the short-time deformation~\cite{Chevillard2008} or assuming Gaussianity of the pressure Hessian \cite{wilczek2014pressure}. These approaches have, for instance, allowed to better understand the Lagrangian evolution of the time-correlations of vorticity alignment \cite{pumir2013tetrahedron,chevillard2011lagrangian}. The other term omitted from the restricted Euler model, the viscous term acts as a damping, and greatly influences the Lagrangian time evolution of the velocity gradient tensor. A popular model to represent the effects of the viscous damping is the linear damping model~\cite{Martin1998a}, which is formally quite simple, but can be regarded as a good approximation for a number of
applications~\cite{Meneveau2011, Martins-Afonso2010}. This is not so at short times, as will be shown in the present investigation.

Whereas, as mentioned above, the evolution of the invariants and their probability density distribution have received a large amount of interest, the Lagrangian time correlations have not been investigated so much. One recent study on the subject is the work by Yu and Meneveau \cite{Yu2010,yu2010scaling}. A tensor-based correlation function was defined to represent the time evolution property along the trajectory of a fluid particle. It was shown that this correlation function always decreases, and the correlation-time is related to the local Kolmogorov time scale. In the present contribution, we will show that a component of this correlation function, the cross correlation, does not always decrease.  Instead, it non-monotonically varies with the time lag, \textit{i.e.} it initially increases for a couple of Kolmogorov time-scales, before it starts to decrease. In order to better understand the short-time evolution of the Lagrangian velocity-gradient correlations, we combine DNS, kinematics and heuristic
modeling.

In Section~II we will define the different components of the Lagrangian velocity gradient correlation tensor and we will show that DNS results predict an increase of one of the components of this tensor.  In Section~III it is rigorously shown that at short times the pressure hessian can cause this effect, neglecting other contributions to the evolution.  A simplified kinematic model will be proposed to explain this interesting phenomenon in Section~IV and  the discussion and conclusion are presented in Section~V.

\section{Behaviour of the Lagrangian velocity gradient correlations in Direct Numerical Simulation}

\subsection{The Lagrangian velocity gradient tensor}

We consider isotropic incompressible turbulence. The Lagrangian velocity gradient, denoted by
\begin{equation}
 A_{ij}(\bm x,t_0|t),
\end{equation}
is the value of $\partial_j u_i$ evaluated at time $t$ at the position of the fluid particle which passed through $\bm x$ at time $t_0$. In the present paper we only consider the case of $t>t_0$. The notation $\partial_i$ is an abbreviation for the partial derivative $\partial/\partial x_i$. We note that in general, for $t\neq t_0$,
\begin{equation}\label{eq:Aij}
A_{ij}(\bm x,t_0|t)\neq \partial_j\left(u_i(\bm x,t_0|t)\right).
\end{equation}
In particular, $A_{ii}(\bm x,t_0|t)=0$ by incompressibility, and this does not hold for $\partial_i u_{i}(\bm x,t_0|t)$. The inequality (\ref{eq:Aij}) complicates the analysis of second-order correlations of the Lagrangian velocity gradient tensor, since we can not link them directly to the velocity-correlations. For brevity, we will denote $A_{ij}(\bm x,t_0|t)$ by $A_{ij}(t)$. Without losing generality, we also define $t_0=0$ in the following sections.

The Lagrangian evolution of $A_{ij}$ is given by
\begin{equation}\label{dotAij}
 \dot{A}_{ij}=-{A}_{pj}{A}_{ip}-P_{ij}+\nu \partial^2_p A_{ij}.
\end{equation}
with $P_{ij}=\partial_i\partial_j p$ the pressure Hessian and $\nu$ the viscosity.

We further introduce the correlation
\begin{equation}
B_{ijmn}(t)=\left<A_{ij}(0)A_{mn}(t)\right>,
\end{equation}
and we omit the parameter $t_0=0$. Since we will only consider statistically stationary flows, the temporal correlations depend only on the time lag $t-t_0=t$.

At $t=t_0$, elementary tensorial kinematics show that, due to isotropy, homogeneity and incompressibility,
\begin{equation}\label{eq:BIJMN}
 B_{ijmn}(0)=\frac{2}{15}\left(\delta_{im}\delta_{jn}-\frac{1}{4}\left(\delta_{ij}\delta_{mn}+\delta_{in}\delta_{jm}\right)\right)\frac{\epsilon}{\nu},
\end{equation}
where the dissipation rate $\epsilon$ is related to the velocity gradients by
\begin{equation}
\epsilon=15\nu\left< \left(\frac{\partial u_1}{\partial x_1}\right)^2\right>.
\end{equation}
All components of  $B_{ijmn}(t_0)$  are thus determined by one scalar invariant, the dissipation. In the following we will consider the time dependence of $B_{ijmn}(t)$, and in particular the transverse auto- and cross-correlations, $B_{1212}$ and $B_{1221}$. Their values at $t=t_0$ are
\begin{equation}\label{eq:BIJMN1}
 B_{1212}(0)=\frac{2}{15}\frac{\epsilon}{\nu},
~~~~~~ B_{1221}(0)=-\frac{1}{30}\frac{\epsilon}{\nu}.
\end{equation}
At later times we will evaluate the different correlations as compared to their value at $t=t_0$. For instance,
\begin{eqnarray}
 \tilde B_{1212}(t)\equiv \frac{B_{1212}(t)}{B_{1212}(0)}.
\end{eqnarray}
Intuitively we would expect the norm of these correlations to decay in time, since dynamics on trajectories have a finite time-correlation. It is observed that at short times this is the case for $\tilde B_{1212}(t)$, %and $\tilde B_{1122}$
but it is not so for $\tilde B_{1221}(t)$.

\subsection{Numerical observations}
\begin{figure}
\begin{center}
{\includegraphics{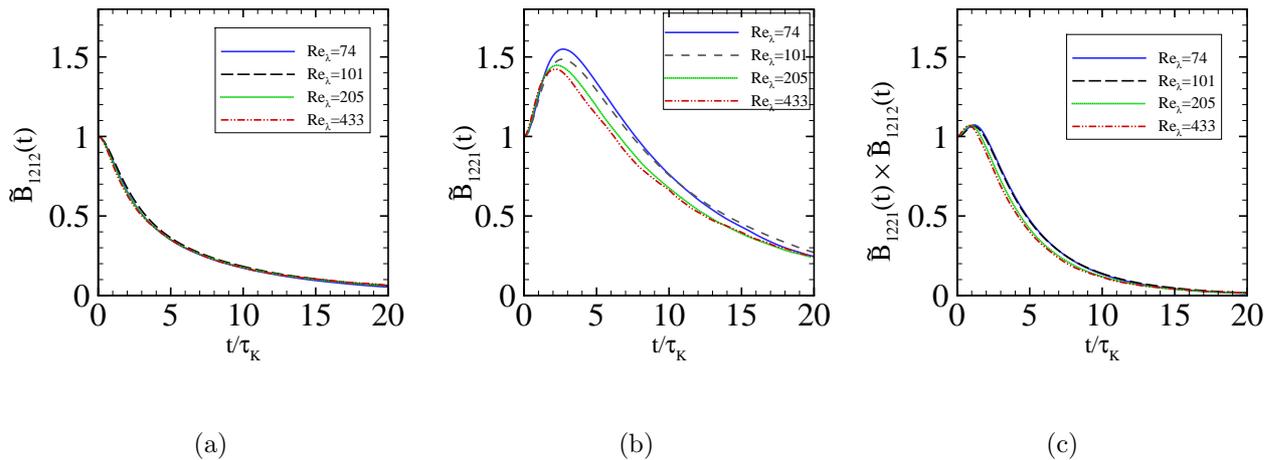}}
  \caption{The two-time correlation functions calculated in DNS databases. The corresponding values at time $t_0=0$ are used for normalization. (a) $\tilde{B}_{1212}(t)$, (b) $\tilde{B}_{1221}(t)$, (c) their product.}
 \label{fig:DNS}
  \end{center}
\end{figure}

A pseudo-spectral method is used to solve the Navier-Stokes equations in a periodically cubic box of size $2 \pi$. A large-scale random forcing scheme is added to the Navier-Stokes equations to produce and maintain statistically stationary isotropic turbulent flows. The details of the calculation can be found in Refs. \cite{Jin2013, He2009a}. The effect of external forcing was discussed by Jeong and Girimaji \cite{Jeong2003}. The isotropic turbulent flows at four Taylor's microscale Reynolds numbers, $Re_{\lambda}=74$, $101$, $205$ and $433$ are simulated to study the effects of the Reynolds number on the cross correlation of velocity gradients. The flow parameters in different flows are listed in Table~\ref{table1}. Here $N$ is the grid resolution in one direction, $u'$ is the root mean square of the fluctuation velocity, $L_f$ is the integral length scale of the flow, $v_K = (\nu \epsilon )^{1/4}$ is the Kolmogorov velocity scale, and $\tau_K = \langle \nu/\epsilon \rangle^{1/2}$ is the Kolmogorov time-scale.  The first three cases are generated using our in-house code and the last one is obtained using the public turbulence database at the Johns Hopkins University\cite{database_Meneveau}. When the turbulent flow field reaches a statistically stationary state, the initial positions of $4\times10^5$ fluid particles are recorded and the trajectories of these particles are then advanced in time using a fourth-order Adams-Moulton method according to ${{d{\bf{x}}} \mathord{\left/ {\vphantom {{d{\bf{x}}} {dt}}} \right. \kern-\nulldelimiterspace} {dt}} = {\bf{u}}({\bf{x}}(t),t)$, where ${\bf{u}}({\bf{x}}(t),t)$ is the fluid velocity experienced by one of the fluid particles, obtained from the Eulerian fluid velocity field using a $6$th-order Lagrangian interpolation~\cite{Yang_He_Wang2008}. The Lagrangian velocity gradient is calculated along the trajectory of a fluid particle. First, we compute the velocity gradient field in an Eulerian frame, then, we obtain the Lagrangian velocity gradient experienced by a fluid particle also using a $6$th-order Lagrangian interpolation. For single-point two-time correlations of the velocity gradient, we obtain the velocity gradient at different times at the fixed initial positions of the $4\times10^5$ fluid particles. As the Lagrangian correlation of velocity gradient decays more slowly than the single-point two-time correlation of the velocity gradient, we calculate the correlations for about $30\tau_K$ so that the Lagrangian correlation of velocity gradient decays to zero. The number of fluid particles used for the determination of the Lagrangian correlation is sufficient to ensure converged, smooth correlation functions, as shown in Fig.~\ref{fig:DNS}.

\begin{table}
\caption{Parameters of the considered flows.}
\begin{ruledtabular}
\begin{tabular}{cccccccccccc}
Case& $N^3$  & $Re_{\lambda}$ &$\epsilon$& $\nu$  &  $u'$ & $L_f$ & $v_{K}$ & $\tau_{K}$ &  \\
I&$128^3$& 74           &3434.7       &0.095   &19.10  &0.91  &4.31&0.0051\\
II&$256^3$&101           &3468.0       &0.049   &19.52  &0.99  &3.62&0.0037 \\
III& $512^3$ &205           &0.2055       &0.001   &0.8722  &3.2283  &0.1197&0.06976\\
IV&$1024^3$&433           &0.0928       &0.000185   &0.681  &1.376  &0.064&0.0446 \\
\end{tabular}
\end{ruledtabular}
\label{table1}
\end{table}

In Fig.~\ref{fig:DNS} it is shown that the correlation $\tilde B_{1212}$
decays monotonically, but this is not so for $\tilde B_{1221}$. This last quantity increases to a value of around $1.5$ after a couple of Kolmogorov times. After that time it starts to decay. At long times this correlation should go to zero, as do the other ones, reflecting the finite correlation time of a turbulent velocity field. The product of $\tilde B_{1221}$ and $\tilde B_{1212}$ also increases at short times and decays at long times. The observed tendencies do not seem to be strongly dependent on the Reynolds number. The evolution of the correlations as a function of time in Kolmogorov-units is roughly the same for all Reynolds numbers, but the influence of the Reynolds number on  $\tilde B_{1221}$ is larger than that on $\tilde B_{1212}$.

\begin{figure}
\begin{center}
  \includegraphics[width=0.4\textwidth]{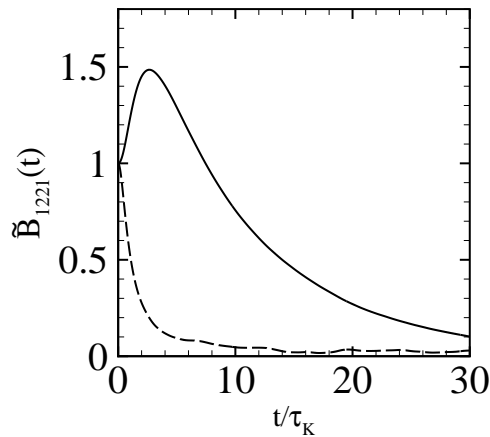}
 \caption{Comparison between the Lagrangian correlation and the single-point two-time correlation of $\tilde B_{1221}(t)$ at Taylor's microscale Reynolds number $Re_{\lambda}=101$, where the solid line denotes the Lagrangian correlation of $\tilde B_{1221}(t)$, and the dashed line the single-point two-time correlation of $\tilde B_{1221}(t)$.}
 \label{fig:Lag_Eur}
  \end{center}
\end{figure}

This short-time increase in $\tilde B_{1221}(t)$ is a typically Lagrangian effect, as illustrated in Fig. \ref{fig:Lag_Eur}, which shows the comparison between the Lagrangian correlation of $\tilde B_{1221}(t)$ and the single-point two-time correlation of $\tilde B_{1221}(t)$, where the latter is the correlation of the velocity-gradient calculated at the same position, \textit{i.e.}, without tracking particles. Evidently, the single-point two-time correlation monotonically decays with time, in contrast to the Lagrangian correlation.

In the next section we will try to explain the different behaviours at short times by kinematic considerations.

\section{Short-time evolution of the correlations}\label{sec:Taylor}

\subsection{Lagrangian correlations}

For short times $(t-t_0)$ we can use the Taylor expansion
\begin{equation}
A^{Taylor}_{ij}(t)=A_{ij}(t_0)+(t-t_0)\dot{A}_{ij}(t_0)+\frac{(t-t_0)^2}{2}\ddot{A}_{ij}(t_0)+\mathcal{O}\left((t-t_0)^3\right).
\end{equation}
The short time evolution of $B_{ijmn}(t)$ can therefore be evaluated by substituting the first non-vanishing contributions of $A^{Taylor}_{ij}(t)$ in $B_{ijmn}(t_0,t)$. We find, up to second order,
\begin{equation}\label{eq:Btaylor}
B_{ijmn}(t)\approx B_{ijmn}(t_0)+(t-t_0)\left<\dot{A}_{ij}(t_0){A}_{mn}(t_0)\right>+\frac{(t-t_0)^2}{2}\left<\ddot{A}_{ij}(t_0){A}_{mn}(t_0)\right>.
\end{equation}
We can evaluate the leading-order term, proportional to $(t-t_0)$, by substituting Eq. (\ref{dotAij}) in Eq. (\ref{eq:Btaylor}). Working out the kinematics for isotropic incompressible turbulence, we find that this term vanishes, reflecting physically the steady-state equilibrium between enstrophy production and dissipation. The vanishing of the first order term can however be shown more directly by considering the time derivative,
\begin{equation}\label{eq:ddtAA}
 \frac{\partial}{\partial t}\left<A_{ij}(t)A_{mn}(t)\right>=0,
\end{equation}
since the flow is statistically stationary.
Therefore, we have immediately,
\begin{equation}\label{eq:}
\left<\dot A_{ij}(t)A_{mn}(t)\right>=-\left<A_{ij}(t)\dot A_{mn}(t)\right>,
\end{equation}
which shows that for $\tilde B_{1212}(t)$
and $\tilde B_{1221}(t)$, the first-order contribution in Eq.~(\ref{eq:Btaylor}) vanishes. In order to explain the short-time behaviour of $\tilde B_{1212}(t)$
and $\tilde B_{1221}(t)$, we therefore need to retain at least the terms in $(t-t_0)^2$.

Using the fact that
\begin{equation}\label{eq::}
 \frac{\partial^2}{\partial t^2}\left<A_{ij}(t)A_{mn}(t)\right>=0,
\end{equation}
we find that
\begin{eqnarray}
 \left<\dot{A}_{ij}(t_0)\dot{A}_{mn}(t_0)\right>= -\frac{1}{2}\left(\left<\ddot{A}_{ij}(t_0){A}_{mn}(t_0)\right>+\left<{A}_{ij}(t_0)\ddot{A}_{mn}(t_0)\right>\right)\nonumber\\
\end{eqnarray}
leading to
\begin{eqnarray}
 \left<\ddot{A}_{12}(t_0){A}_{12}(t_0)\right>= -\left<\dot{A}_{12}(t_0)\dot{A}_{12}(t_0)\right>,\nonumber\\
 \left<\ddot{A}_{12}(t_0){A}_{21}(t_0)\right>= -\left<\dot{A}_{12}(t_0)\dot{A}_{21}(t_0)\right>,
\end{eqnarray}
which helps us to get rid of the second-order time derivatives. Isotropy allows to express
\begin{equation}\label{dotAdotA}
 \left<\dot{A}_{ij}(t_0)\dot{A}_{mn}(t_0)\right>=\frac{1}{30}\left(\frac{\epsilon}{\nu}\right)^2\left[(4F-G)\delta_{im}\delta_{jn}-(F+G)\delta_{ij}\delta_{mn}+(4G-F)\delta_{in}\delta_{jm}\right],
\end{equation}
with
\begin{eqnarray}
F=\left<\dot{A}_{ij}\dot{A}_{ij}\right>\left(\frac{\nu}{\epsilon}\right)^2, ~~~~G=\left<\dot{A}_{ij}\dot{A}_{ji}\right>\left(\frac{\nu}{\epsilon}\right)^2.
\end{eqnarray}
Using expressions (\ref{eq:BIJMN}), (\ref{eq:Btaylor}) and (\ref{dotAdotA}), we find for the normalized correlations,
\begin{eqnarray}\label{eq:B123}
 \tilde B_{1212}(t)\approx 1-\frac{1}{8}\left(\frac{t-t_0}{\tau_K}\right)^2(4F-G),\\
% \tilde B_{1122}(t_0,t)\approx 1-\frac{1}{2}\left(\frac{t-t_0}{\tau_K}\right)^2(F+G)\\
 \label{eq:B123b}\tilde B_{1221}(t)\approx 1+\frac{1}{2}\left(\frac{t-t_0}{\tau_K}\right)^2(4G-F).
\end{eqnarray}
This shows that if we can determine the scalar quantities $F$ and $G$, we can determine all the initial trends of the Lagrangian correlation functions. We will first see which constraints are implied by purely kinematic constraints, \textit{i.e.}, without introducing the Navier-Stokes equations. Auto-correlations in stationary turbulence can be assumed to be decaying functions of time, so that, using this constraint  on $B_{1111}$ and $B_{1212}$, we have $-F\le G\le 4F$, and $0\le F$. This allows but does not demonstrate that $\tilde B_{1221}(t)$ is an increasing function in time, since for that, we should have $F/4\le G$. It is normal that at this point we cannot demonstrate this, since we have not used any information on the Navier-Stokes equations, only considerations assuming isotropy and incompressibility.

Let us at this point add such dynamic information, step by step. For instance consider that only the viscous term on the RHS of (\ref{dotAij}) is non-zero. In that case we find, by isotropy that $G=0$, and both correlations in Eq.~(\ref{eq:B123}) and Eq.~(\ref{eq:B123b}) initially decrease. If we now consider a different case, where only the pressure Hessian is non-zero, we find $F=G$ since the pressure Hessian is symmetric in its indices $P_{ij}=P_{ji}$. In that case the cross-correlation increases whereas the auto-correlation decreases, as observed in the simulations. This does not prove that it is the pressure Hessian alone which is responsible for the increasing correlation, the self-interaction term (first term on the RHS of Eq.~(\ref{dotAij})) can also play a role. For that term we have not succeeded to show any simple symmetry properties in a rigorous way. We have therefore proceeded to measure $G/F$ in the direct numerical simulations and we found that  $G/F \approx 0.6$ in the simulations, roughly independent of the
Reynolds number. This is in the interval $1/4\le G/F \le 4$, in which auto-correlations decrease and  $\tilde B_{1221}(t)$  increases in time.

\subsection{Single-point two-time correlation}

\begin{figure}
\begin{center}
  \includegraphics{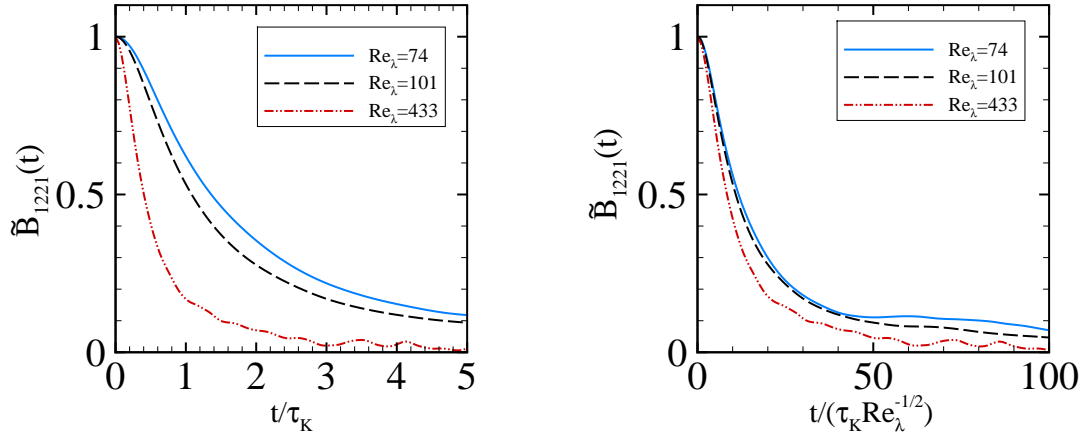}~
  \caption{Single-point two-time correlation of the gradient cross-correlation $\tilde B_{1212}(t)$ at different Reynolds numbers, scaled by Kolmogorov time-units (left) and sweeping time-units (right).}
 \label{fig:ScaleEulerian}
  \end{center}
\end{figure}

For the case of the single-point two-time correlations of the velocity gradient tensor, we have an additional term on the RHS of the velocity-gradient evolution equation,
\begin{equation}\label{dotAijE}
 \partial_t A_{ij}=-u_m\partial_mA_{ij}-{A}_{pj}{A}_{ip}-P_{ij}+\nu \partial^2_p A_{ij}.
\end{equation}
the advection term $-u_m\partial_mA_{ij}$ should now be taken into account in the evaluation of $F$ and $G$. Doing so, ignoring all the other terms, we have
\begin{equation}
\left<\dot{A}_{ij}\dot{A}_{ij}\right>_S=\left<u_m\partial_m A_{ij}u_n\partial_nA_{ij}\right>.
\end{equation}
with the index $S$ denoting sweeping. We assume that the sweeping velocity and the velocity gradients are roughly independent (see for instance \cite{He2006, Zhao2009}), so that
\begin{equation}
\left<\dot{A}_{ij}\dot{A}_{ij}\right>_S=\left<u_m u_n\right>\left<\partial_m A_{ij}\partial_nA_{ij}\right>.
\end{equation}
Invoking isotropy for the  large-scale sweeping, this relation becomes
\begin{equation}
\left<\dot{A}_{ij}\dot{A}_{ij}\right>_S=\frac{1}{3}\left<u_m u_m\right>\left<\partial_n A_{ij}\partial_nA_{ij}\right>.
\end{equation}
This term can be expressed as a function of the energy spectrum. Assuming Kolmogorov scaling for this spectrum,
\begin{equation}\label{eq:Ek}
 E(k)=\nu^{5/4} \epsilon^{1/4}f_u(k\eta),
\end{equation}
we find that
\begin{equation}
F_S\equiv\left<\dot{A}_{ij}\dot{A}_{ij}\right>_S\left(\frac{\nu}{\epsilon}\right)^2=\frac{2}{3}\frac{\left<u_m u_m\right>}{(\epsilon\nu)^{1/2}}\int \zeta^4 f(\zeta) d\zeta,
\end{equation}
where $C_\zeta=\int \zeta^4 f(\zeta) d\zeta$ is supposed to be of order unity. We similarly find  $G_S=0$. If we substitute this in the Taylor expansion, we find that
\begin{eqnarray}\label{eq:B123a}
 \tilde B_{1212}(t_0,t)=\tilde B_{1122}(t_0,t)=\tilde B_{1221}(t_0,t)\approx 1-\frac{1}{3}C_\zeta\left(\frac{t-t_0}{\tau_K}\right)^2 \frac{\left<u_m u_m\right>}{(\epsilon\nu)^{1/2}}.
\end{eqnarray}
This shows that sweeping should decorrelate all correlations on the same time-scale, and that this decorrelation is not independent of Reynolds number when scaled by Kolmogorov time-units. If one would like to collapse the different time-correlations, one should scale them by time units $\mathcal T=\tau_K R_\lambda^{-1/2}$. In Fig. \ref{fig:ScaleEulerian} we show that this scaling allows to collapse the data better than the scaling by Kolmogorov-units.
For high Reynolds numbers this decorrelation should therefore be much faster and the other terms, which were observed to lead to decorrelations over several Kolmogorov time-scales, are therefore most probably negligible.

In order to obtain some more intuition on the dynamics, introducing the influence of the flow topology, we will now discuss simple heuristic models for the Lagrangian dynamics.

\section{Heuristic models for the Lagrangian velocity gradient correlations}

\subsection{Strain and vorticity correlations}\label{sec:SOmega}

The velocity gradient tensor can be decomposed without loss of generality into its rotation and strain part. This so-called Helmholtz decomposition (HD) leads to
\begin{equation}
A_{ij} = S_{ij} + \Omega_{ij}
\end{equation}
with $S_{ij}=(\partial_j u_i + \partial_i u_j)/2$ and $\Omega_{ij}=(\partial_j u_i - \partial_i u_j)/2$. If we assume that both contributions remain independent of each other over the considered time-interval, we have

\begin{equation}
\begin{split}
B_{ijmn}(t)= & \left\langle (S_{ij}(0) + \Omega_{ij}(0)) (S_{mn}(t) + \Omega_{mn}(t)) \right\rangle \\
\approx & \left\langle S_{ij}(0)S_{mn}(t)\right\rangle + \left\langle \Omega_{ij}(0)\Omega_{mn}(t)) \right\rangle. \label{eq:BijmnPumir}
\end{split}
\end{equation}
In several recent studies of Lagrangian turbulence it was observed that the time-correlations of strain and rotation could be approximated by exponentially decaying correlation functions \cite{Shin2005, Vincenzi2007, Pumir2011}. The two quantities are however correlated over different timescales. The characteristic time scales of shear and rotation are, respectively, about $2.3\tau_K$ and $7.2\tau_K$ with $\tau_K$ the Kolmogorov timescale. Combining this information with expression (\ref{eq:BijmnPumir}),  we find
\begin{equation}
\begin{split}
\tilde{B}_{1212}(t) = & \dfrac{3}{8}\exp\left(-\dfrac{t}{2.3\tau_K}\right)+\dfrac{5}{8}\exp\left(-\dfrac{t}{7.2\tau_K}\right), \\
\tilde{B}_{1221}(t) = & -\dfrac{3}{2}\exp\left(-\dfrac{t}{2.3\tau_K}\right)+\dfrac{5}{2}\exp\left(-\dfrac{t}{7.2\tau_K}\right).
\end{split}
\end{equation}

\begin{figure}
\begin{center}
{\includegraphics{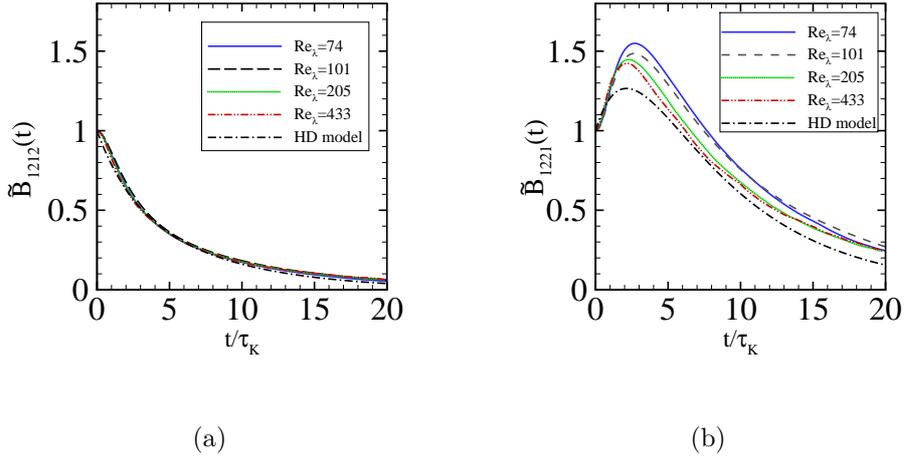}}~
\caption{Comparison between the HD model with DNS data.With increasing Reynolds numbers, the cross correlations $\tilde{B}_{1221}(t)$ from DNS data gradually approach the HD model. (a) $\tilde{B}_{1212}(t)$, (b) $\tilde{B}_{1221}(t)$.}
\label{HDM_VS_DNS}
\end{center}
\end{figure}

These expressions, termed as  HD model, are compared to the DNS data, as shown in Fig.~\ref{HDM_VS_DNS}. It shows that the HD model captures the main characteristics for the Lagrangian correlations of the velocity gradients. However, the HD model does not reproduce the correct peak value of $\tilde{B}_{1221}$.

An additional insight is obtained by considering the two-dimensional case. In this case the vorticity is purely advected,
\begin{equation}
 \dot{\omega}=\nu \Delta \omega.
\end{equation}
On a Lagrangian trajectory, the vorticity is thus only decorrelated by viscous diffusion of vorticity. We write the vorticity as a function of the velocity gradients, $\omega=(A_{21}-A_{12})$, so that the Lagrangian autocorrelation writes,
\begin{equation}
 B_\omega(t)=\frac{\left<\omega(t_0)\omega(t)\right>}{\left<\omega(t_0)\omega(t_0)\right>}=2\frac{\left<A_{12}(t_0)A_{12}(t)\right>-\left<A_{12}(t_0)A_{21}(t)\right>}{\left<\omega(t_0)\omega(t_0)\right>}.
\end{equation}
Since in two dimensions $\left<A_{12}(t_0)A_{21}(t_0)\right>=-\left<A_{12}(t_0)A_{12}(t_0)\right>/3$, we find that,
\begin{equation}
 B_\omega(t)=\frac{3}{4}\tilde B_{1212}(t)+\frac{1}{4}\tilde B_{1221}(t).
\end{equation}
If we neglect the viscous diffusion, the vorticity will not decorrelate on a trajectory. In this case  $B_\omega(t)=1$ and we find,
\begin{equation}
\tilde B_{1221}(t)=4-3\tilde B_{1212}(t).
\end{equation}
This shows that if the autocorrelation $\tilde B_{1212}(t)$ decreases, say, as $\tilde B_{1212}(t)=1-((t-t_0)/\tau)^2$, the cross-correlation will increase as
\begin{equation}
\tilde B_{1221}(t)=1+3((t-t_0)/\tau)^2.
\end{equation}
In two dimensions the increase of the correlation $\tilde B_{1221}(t)$ can thus be understood by the inviscid mechanism of vorticity advection. The longer correlation time of the vorticity as compared to strain, in three dimensions could be reminiscent of this mechanism, even though in three dimensions the vorticity is not conserved due to the presence of vortex stretching. The investigation of the lagrangian time-correlations of the velocity-gradient in two-dimensional turbulence is left for further research.

\subsection{Lagrangian correlations by assuming locally constant velocity gradients}

We will here try to identify the different effects that lead to the observed results. Therefore we will consider a given velocity field with locally constant velocity gradients. Subsequently we will show how, following a fluid particle on a trajectory, rotation and shear influence the Lagrangian correlations. Subsequently we add damping and as a last feature we allow the initially considered velocity field to decorrelate itself in time. This step by step complexification allows to disentangle the different physical mechanisms leading to the observations in the DNS.

\subsubsection{Stretching and rotating in an inviscid field}

We here introduce a simple heuristic model, where we can analytically compute the correlations. Our starting point is a velocity gradient field which we assume to be locally uniform and constant. In that case, locally the field can be considered as a 2D state (\textit{c.f.}, section 2.3.2 of Ref. \cite{Sagaut_Cambon_Book}). We choose the coordinate-system such that the velocity gradient is in the $x,y$-plane, with its axes chosen such that
\begin{equation}\label{eq:MatrixA}
\bm{A} = \left[
     \begin{array}{lll}
      -S & - \Omega& 0\\
       \Omega & S&0\\
       0 & 0&0\\
     \end{array}
    \right]
\end{equation}
with $S$ the pure irrotational strain rate and $\Omega$ the angular rotation rate. The associated flow-field is obviously not isotropic, but we will consider that the whole space is filled with an infinite number of local structures, and for each structure the characteristic orientations of the velocity gradient tensor are randomly distributed. The averages over all orientations will yield us isotropic statistics.

A rotation in a 3D space can be expressed using the rotation-transform tensor,
\begin{equation}
\bm{Q}_E^{(\theta, \beta, \gamma)} = \left[
     \begin{array}{lll}
      \cos\theta\cos\gamma-\cos\beta\sin\theta\sin\gamma & \sin\theta\cos\gamma+\cos\beta\cos\theta\sin\gamma & \sin\beta\sin\gamma \\
      -\cos\theta\sin\gamma-\cos\beta\sin\theta\cos\gamma & -\sin\theta\sin\gamma+\cos\beta\cos\theta\cos\gamma & \sin\beta\cos\gamma \\
      \sin\beta\sin\theta & -\sin\beta\cos\theta & \cos\beta
     \end{array}
    \right],
\end{equation}
where $\theta$, $\beta$ and $\gamma$ are a group of Euler angles in the directions of $z$, $x$ and $z$ respectively. In particular, when $\beta=\gamma=0$ it yields a rotation in the $x,y$-plane with angle $\theta$:
\begin{equation}
\bm{Q}^{(\theta)} = \left[
     \begin{array}{lll}
      \cos\theta & \sin\theta &0 \\
      -\sin\theta & \cos\theta&0 \\
      0 & 0&1
     \end{array}
    \right].
\end{equation}
Rotating a tensor in 3D yields the velocity gradient
 \begin{equation}
\bm{A}^{(\theta, \beta, \gamma)} = \bm{Q}_E^{(\theta, \beta, \gamma)^T}\bm{A}\bm{Q}_{E}^{(\theta, \beta, \gamma)},
\end{equation}
with the superscripts $(\theta, \beta, \gamma)$ denoting the rotation angle, and $T$ the transposition of the matrix. This yields the off-diagonal components of the velocity gradient tensor $A_{12}^{(\theta, \beta, \gamma)}$ and $A_{21}^{(\theta, \beta, \gamma)}$.

The single-time velocity-gradient correlations $B_{1212}$ and $B_{1221}$ over an infinite number of randomly oriented realizations can then be calculated using the Haar measure of Euler angles:
\begin{equation}\label{eq:BSW}
\begin{split}
B_{1212}\equiv\langle A_{12}A_{12}\rangle = & \dfrac{1}{8\pi^2}\int_0^{2\pi}d\theta \int_0^{\pi}\sin\beta d\beta \int_0^{2\pi}d\gamma A_{12}^{(\theta, \beta, \gamma)}A_{12}^{(\theta, \beta, \gamma)}  = \dfrac{1}{15}(3S^2+5\Omega^2), \\
B_{1221}\equiv\langle A_{12}A_{21}\rangle = & \dfrac{1}{8\pi^2}\int_0^{2\pi}d\theta \int_0^{\pi}\sin\beta d\beta \int_0^{2\pi}d\gamma A_{12}^{(\theta, \beta, \gamma)}A_{21}^{(\theta, \beta, \gamma)}  = \dfrac{1}{15}(3S^2-5\Omega^2). \\
\end{split}
\end{equation}
In three dimensions Eq. (\ref{eq:BIJMN1}) shows that $B_{1221}=-B_{1212}/4$. Combining this with Eq.~(\ref{eq:BSW}) we find that $S^2=\Omega^2$, $B_{1212}=8S^2/15$ and $B_{1221}=-2S^2/15$. Comparing to Eq.~(\ref{eq:BIJMN1}) we also have $S^2 = \epsilon/(4\nu)$ which is in agreement with Eq.~(\ref{eq:MatrixA}) and the definition of the dissipation rate.

We will now compute the evolution of the Lagrangian velocity-gradient correlations. In our homogeneous velocity-gradient field, we follow a fluid particle and we will determine how the velocity gradient is modified through the interaction of strain and rotation. In Section III it was shown that the pressure Hessian might be an important factor for the short-time phenomenon of Lagrangian correlation, thus an appropriate two-time model should consider the pressure Hessian. As it is difficult to rigorously take account of the pressure effect, we borrow the assumption of ``frozen velocity gradient field" by Chevillard \textit{et al.} \cite{Chevillard2008, Chevillard2006PRL}. Under this assumption the deformation of a fluid particle is driven by a constant velocity gradient field during a short time of the order $\tau_K$. This deformation, on the one hand implies the effect of pressure Hessian \cite{Chevillard2008}, on the other hand unfreezes and changes the velocity gradient tensor. Specifically, for the present
model, when an appropriate coordinate system is chosen, from Eq. (\ref{eq:MatrixA}) the influence of this frozen velocity gradient field is divided into the effects of stretching and rotation.

\begin{figure}
\begin{center}
{\includegraphics{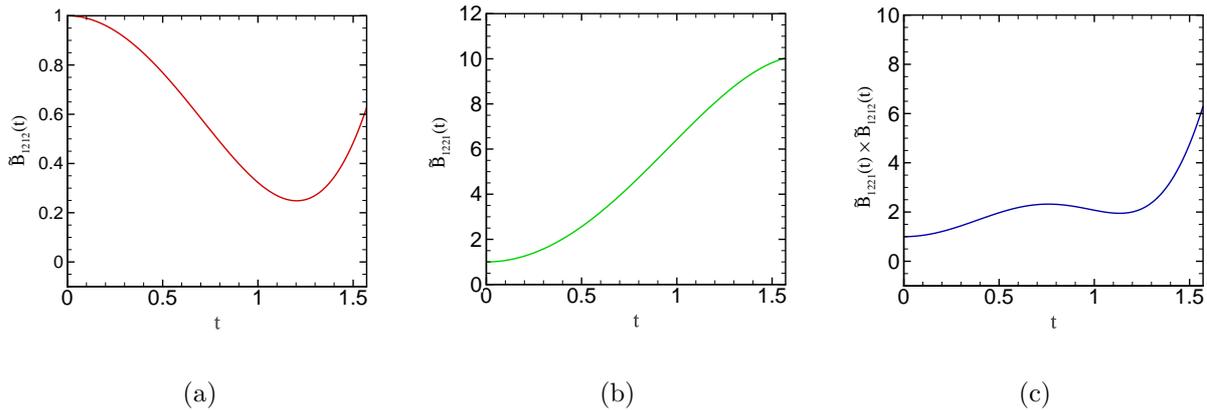}}~
\caption{The two-time correlation functions in a uniform stationary velocity gradient field without damping. (a) $\tilde{B}_{1212}(t)$, (b) $\tilde{B}_{1221}(t)$, (c) their product.}
\label{fig:nodiss}
\end{center}
\end{figure}

Considering the stretching effect by the uniform strain field, applying the strain in matrix Eq.~(\ref{eq:MatrixA}) on a velocity gradient matrix $\bm{A}$ with a given time $t$ and shear rate $S$ leads to
\begin{equation}
\bm{A}^{S}(t) = \bm{A}
\left[
\begin{array}{lll}
e^{-St} & 0& 0 \\
0 & e^{St}& 0 \\
0 & 0& 1
\end{array}
\right].
\end{equation}
Then, by rotating this matrix over $\phi=\Omega t$, the gradient tensor is

\begin{equation}
\bm{A}^{S,(\phi)}(t) = \bm{Q}^{(\phi)^T}\bm{A}^{S}(t)\bm{Q}^{(\phi)}.
\end{equation}
Note that here the rotation is always in the $x,y$-plane, and the transform matrix $\bm{Q}$, rather than $\bm{Q}_E$, should be used. Similar to the previous subsection, considering the random orientations, we have
\begin{equation}
\bm{A}^{S,(\phi), (\theta, \beta, \gamma)}(t) = \bm{Q}_E^{(\theta, \beta, \gamma)^T}\bm{Q}^{(\phi)^T}\bm{A}^{S}(t)\bm{Q}^{(\phi)}\bm{Q}_E^{(\theta, \beta, \gamma)},
\end{equation}

and the Lagrangian two-time correlation $B_{1212}(t)$ and $B_{1221}(t)$ are then
\begin{equation}\label{AA_nodiss}
\begin{split}
B_{1212}(t) = & \dfrac{1}{8\pi^2}\int_0^{2\pi}d\theta \int_0^{\pi}\sin\beta d\beta \int_0^{2\pi}d\gamma A_{12}^{(\theta, \beta, \gamma)}A_{12}^{S, (\phi), (\theta, \beta, \gamma)}, \\
B_{1221}(t) = & \dfrac{1}{8\pi^2}\int_0^{2\pi}d\theta \int_0^{\pi}\sin\beta d\beta \int_0^{2\pi}d\gamma A_{12}^{(\theta, \beta, \gamma)}A_{21}^{S, (\phi), (\theta, \beta, \gamma)}.
\end{split}
\end{equation}
From Eqs. (\ref{AA_nodiss}) and (\ref{eq:BSW}), the normalized values can be obtained as
\begin{equation}\label{AA_nodiss_final}
\begin{split}
&\tilde{B}_{1212}(t) \equiv \frac{B_{1212}(t)}{B_{1212}(0)} = \frac{1}{3S^2 + 5 \Omega^2}\left(\left( 5\Omega^2 + 3 S^2 \cos(2\phi) \right) \cosh(S t) - 3 S\Omega \sin(2\phi) \sinh(S t)\right),\\
&\tilde{B}_{1221}(t) \equiv \frac{B_{1221}(t)}{B_{1212}(0)} = \frac{1}{5 \Omega^2 - 3 S^2}\left(\left( 5\Omega^2 - 3 S^2 \cos(2\phi) \right) \cosh(S t) + 3 S\Omega \sin(2\phi) \sinh(S t)\right).
\end{split}
\end{equation}
Using the isotropic result $S=\Omega$, these correlations can be expressed as a function of $\Omega$ alone. If we make the link with a turbulent flow, the dominant rotation time-scale will be of the order of the Kolmogorov timescale. We therefore define $\Omega^{-1}$ for normalization with $\Omega^{-1} \sim \tau_K$. The resulting two-time correlation functions are shown in Fig. \ref{fig:nodiss}. It is shown that at short time, $\tilde{B}_{1212}(t)$ decreases while $\tilde{B}_{1221}(t)$ increases. Also the product of these two correlation functions, shown in Fig. \ref{fig:nodiss}(c), increases. Indeed, the phenomenological picture that we obtain from these considerations is the following: a fluid particle that is moving in vortical motion will change its orientation. The rotation of the local velocity gradient will then reorient so that after a typical small scale turn-over time the velocity-gradient $\partial_1 u_2$ will have changed towards a $\partial_2 u_1$ local gradient. Apparently, this correlation increases from its initial value if we consider a given velocity-gradient field. At long times all correlations attain nonphysically high values. Indeed, at long-times the correlations are expected to decrease due to turbulent and viscous diffusion. We will therefore add a damping to the present model.

\subsubsection{Adding a damping function to the correlations}

In order to improve the temporal behaviour of the velocity-gradient correlation model, we add a damping function
\begin{equation}\label{AA_diss_c1}
\begin{split}
\tilde B_{1212}(t) = & f(t)\left(\frac{1}{3S^2 + 5 \Omega^2}\left(\left( 5\Omega^2 + 3 S^2 \cos(2\phi) \right) \cosh(S t) - 3 S\Omega \sin(2\phi) \sinh(S t)\right)\right), \\
\tilde B_{1221}(t) = & f(t)\left(\frac{1}{5 \Omega^2 - 3 S^2}\left(\left( 5\Omega^2 - 3 S^2 \cos(2\phi) \right) \cosh(S t) + 3 S\Omega \sin(2\phi) \sinh(S t)\right)\right).
\end{split}
\end{equation}
Traditional damping models usually assume linear dissipation \cite{Martin1998a}, which is probably a good approximation for $t\gg \tau_K$. However, we have seen  that rigorously, for $t\ll \tau_K$ (expression (\ref{eq:B123})), the damping should be a quadratically decaying function of time. We have tried both types of damping. For the linear damping we use $f(t) = \exp(-c_1\Omega t)$ and for the quadratic damping $f(t) = \exp(-c_2(\Omega t)^2)$, with $c_1=c_2=2.2$, a constant which is arbitrarily chosen, but the value of which not qualitatively changes the behaviour as long as it is of order unity.

\begin{figure}
\begin{center}
{\includegraphics{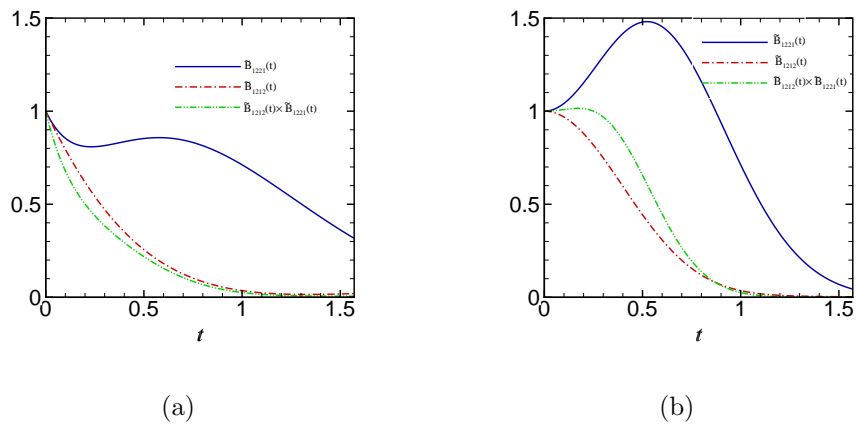}}~
  \caption{The two-time correlation functions with damping. Red line: $\tilde{B}_{1212}(t)$; blue line: $\tilde{B}_{1221}(t)$; green line: their product. (a) Linear damping function; (b) quadratic damping function.}
 \label{fig:AA_diss}
  \end{center}
\end{figure}
It is shown in Fig.~\ref{fig:AA_diss} that the additional damping term allows the correlations to tend to zero for long times, as expected. However, the initial increase of the cross-correlation is only observed in the case of a quadratic damping. The linear damping is incompatible with this feature.  The peak time is related to the characteristic time of the damping model (\textit{c.f.} Refs. \cite{Martin1997, Martins-Afonso2010}). This shows that if models for the Lagrangian evolution of the velocity-gradient tensor are to produce the short time statistics correctly, a quadratic damping should be used. Indeed, comparing to DNS results, for longer times the quadratic damping is over dissipative and the linear damping might prove to be more physically adapted. Hence, an interpolation between the two behaviours should be used in practice. Another possibility is to introduce more sophisticated damping models where the damping-timescale is not constant, but evolves over its Lagrangian trajectory \cite{Jeong2003}. We have instead chosen to improve the damping by evolving the quantities $S$ and $\Omega$ in our phenomenological model, using the timescales discussed in Sec. \ref{sec:SOmega}.

\subsubsection{Damping the correlations using multiple timescales}

\begin{figure}
\begin{center}
{\includegraphics{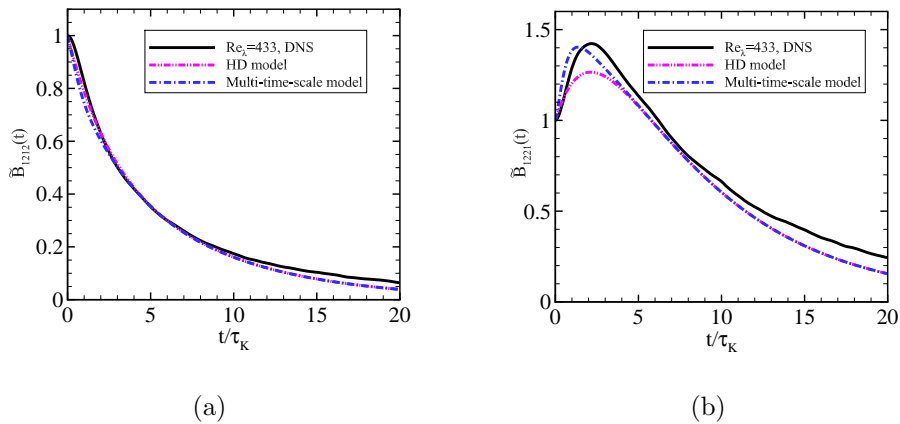}}~
\caption{Comparison between the multi-time-scale kinetic model with HD model and DNS data, where $c_r = 9.0$ in the multi-time-scale kinetic model. (a) $\tilde{B}_{1212}(t)$, (b) $\tilde{B}_{1221}(t)$.}
\label{MTS_VS_DNS}
\end{center}
\end{figure}

An alternative to the ad-hoc damping function $f(t)$ is to allow the velocity gradients to decorrelate in the inviscid model (\ref{AA_nodiss_final}), taking into account that $S_{ij}$ and $\Omega_{ij}$ are governed by different time scales (as in section \ref{sec:SOmega} and discussed in references \cite{Shin2005, Vincenzi2007, Pumir2011}). Additional coefficients are also required to relax the assumption of a frozen velocity gradient field and will be explained later. The multi-time-scale kinetic model then writes
\begin{equation}\label{m-t-s model}
\begin{split}
\tilde{B}_{1212}(t) = \frac{1}{3S(0)^2 + 5 \Omega(0)^2} & \left( \left( 5\Omega(t)^2 + 3 S(t)^2 \cos(2t\Omega_r(t)) \right) \cosh(tS_r(t)) \right.\\
&\left. - 3 S(t)\Omega(t) \sin(2t\Omega_r(t)) \sinh(tS_r(t))\right),\\
\tilde{B}_{1221}(t) = \frac{1}{5 \Omega(0)^2 - 3 S(0)^2} & \left( \left( 5\Omega(t)^2 - 3 S(t)^2 \cos(2t\Omega_r(t)) \right) \cosh(tS_r(t)) \right.\\
&\left. + 3 S(t)\Omega(t) \sin(2t\Omega_r(t)) \sinh(tS_r(t))\right),
\end{split}
\end{equation}
where $S(t)=S(0) \exp(- t/(4.6\tau_K))$, $\Omega(t)=\Omega(0) \exp(- t/(14.4\tau_K))$, $S_r(t) = S(c_r t)$ and $\Omega_r(t) = \Omega(c_r t)$. The coefficient $c_r$ is the relaxation of the assumption of frozen velocity gradient field, which allows different damping rates between the transform field and the defrozen quantity field. In practice if we choose $c_r = 9.0$, a good agreement with DNS results can be obtained (see Fig.~\ref{MTS_VS_DNS}). We remark here that the assumption of frozen velocity gradient field by Chevillard \textit{et al.} \cite{Chevillard2008, Chevillard2006PRL}, does not mean a constant field, but implies a delay effect between the transform and the defrozen quantity field. According to Chevillard \textit{et al.}, this delay is physical and corresponds to the pressure redistribution effects. In the multi-time-scale model, the coefficient $c_r$ thus relaxes the freezing assumption and quantitatively describes this phenomenon of delay.

\subsubsection{Assessment of the models and comparison to DNS}

We have thus reproduced the observations in the simulations using heuristic models. The agreement of the multi-time-scale model with DNS is better than that of the Helmholtz-decomposition assuming exponential time-correlations for strain and vorticity. However, the level of sophistication of the multi-time-scale model is significantly higher and we have added an extra model-constant, characterizing the different damping rates between the transform field and the defrozen quantity field. Rather than a practical model, the multi-time-scale model should be considered as a way to disentangle the different effects. Indeed, it shows that the initial tendencies of the velocity-gradient correlations can be reproduced by the combined effects of translation and rotation. At long times either a Gaussian/exponential damping function should be added to the correlation, or the vorticity and strain along the trajectory should be taken to be decreasing functions in time. Finally, to optimize the agreement with DNS, the initially frozen velocity field should be defrozen by introducing a coefficient of relaxation. This model extends the studies of Chevillard \textit{et al.} \cite{Chevillard2008, Chevillard2006PRL} and proposes a relaxation of the assumption of a frozen velocity field.

\section{Discussion and conclusion}\label{sec:discussion}

The fact observed in this paper, \textit{i.e}., both $\tilde{B}_{1221}(t)$ and $\tilde{B}_{1212}(t)\tilde{B}_{1221}(t)$ always increase at short time, is surprising because these kinds of correlation are usually decaying in dissipative systems. We have tried to understand the short-time evolution of these correlations using kinematical considerations. In section \ref{sec:Taylor} we have attempted to obtain rigorous results for the short-time correlations. It was shown, without even considering the Navier-Stokes equations, that the short-time evolution of the Lagrangian velocity gradient correlations is entirely determined by two invariants,  $F\sim\left<\dot{A}_{ij}\dot{A}_{ij}\right>$ and $G\sim\left<\dot{A}_{ij}\dot{A}_{ji}\right>$. An increase of the Lagrangian cross-correlation at short times should be observed if the ratio of the two invariants lies in the interval $1/4\le G/F \le 4$.  Subsequently we showed that the influence of viscosity gives $G/F = 0$, thereby damping all correlations, as expected. However,
symmetry arguments show that if we only consider the pressure Hessian, we find  $G/F = 1$, which should give rise to an increase of the cross-correlation. We measured the ratio and found that $G/F \approx 0.6$ so that the initial evolution cannot be explained uniquely by the influence of the pressure Hessian. Also the influence of damping and self-interaction should be considered to understand the full picture.

Considering the Lagrangian time-correlation of the vorticity, it was shown that in the two-dimensional inviscid case the increases of $\tilde{B}_{1212}(t)$ and $\tilde{B}_{1221}(t)$ are kinematically coupled and an increase of one of the correlations implies the decorrelation of the other. The presence of vortex stretching in three dimensions does not allow such a simple conclusion, but if the vorticity correlation is more persistent than the $\tilde{B}_{1212}(t)$ correlation, an initial increase of $\tilde{B}_{1221}(t)$ is expected. By contrast, the presence of sweeping in the Eulerian framework does not allow to observe any increase of time-correlations.

To give a more intuitive understanding of the link between flow-topology and Lagrangian evolution, two heuristic models were proposed. The Helmholtz-decomposition model can qualitatively reproduce the observed effect. Then we aim at disentangling the different physical mechanisms leading to the observations in the DNS, which finally lead to the multi-time-scale model validated by the quantitative agreement with DNS. These models, on the one hand, explain the physical roles of stretching, rotation, pressure and damping in the present observation; on the other hand, show limitations of the traditional linear damping at short times, and support a multiple-timescale damping to relax the assumption of a frozen velocity gradient field.

\section*{Acknowledgement}
This work is supported by $973$ program of China (2013CB834100), the National Natural Science Foundation of China (11202013, 11572025, 51420105008, 11232011 and 11472277),the National Natural Science Associate Foundation (NSAF) of China (grant number U1230126). GDJ benefitted from the hospitality of the Nordic Institute for Theoretical Physics under the auspices of the program ``Dynamics of Particles in Flows: Fundamentals and Applications" in June 2014 in Sweden. The anonymous referees are much appreciated for the constructive suggestions to improve the manuscript.
\bibliography{Subgrid}
\bibliographystyle{unsrt}

\end{document}